\documentclass[prd,onecolumn,nopacs,nokeys]{revtex4}
\usepackage{amsfonts}
\usepackage{amsmath}
\usepackage{amssymb}
\usepackage{bbm}
\usepackage[utf8]{inputenc}
\usepackage{tensor}
\usepackage{slashed}
\usepackage[centertableaux ]{ytableau}
\usepackage{hyperref}
\usepackage{natbib}
\usepackage{enumerate}
\usepackage{braket,mleftright}
\usepackage{graphicx}
\usepackage[caption=false]{subfig}
\usepackage{subfig}
\usepackage{cleveref}

\begin{document}

\title{Cosmic-ray antiproton excess from annihilating tensor dark matter.}
\author{H. Hern\'{a}ndez-Arellano\footnote{email:h.hernandezarellano@ugto.mx, corresponding author}} 
\address{Departamento de F\'{\i}sica, Universidad de
  Guanajuato, Lomas del Bosque 103, Fraccionamiento Lomas del
  Campestre, 37150, Le\'{o}n, Guanajuato, M\'{e}xico. }
\author{M. Napsuciale\footnote{email:mauro@fisica.ugto.mx}} 
\address{Departamento de F\'{\i}sica, Universidad de
  Guanajuato, Lomas del Bosque 103, Fraccionamiento Lomas del
  Campestre, 37150, Le\'{o}n, Guanajuato, M\'{e}xico. }
\author{S. Rodr\'{\i}guez\footnote{email: simonrodriguez@uadec.edu.mx}}
\address{Facultad de Ciencias F\'isico-Matem\'aticas,
  Universidad Aut\'onoma de Coahuila, Edificio A, Unidad
  Camporredondo, 25000, Saltillo, Coahuila, M\'exico.}
 \author{R. Ram\'{\i}rez-Guzm\'{a}n \footnote{email:or.ramirezguzman@ugto.mx}}
 \address{Departamento de F\'{\i}sica, Universidad de
  Guanajuato, Lomas del Bosque 103, Fraccionamiento Lomas del
  Campestre, 37150, Le\'{o}n, Guanajuato, M\'{e}xico. }

\begin{abstract}
In this paper we calculate the antiproton production in the annihilation of tensor dark matter and explore the possibility that the excess of antiprotons in the range $E_{K}=10-20 ~GeV$ 
reported by several groups in the analysis of the AMS-02 Collaboration data is due to this production mechanism. 
We find that these contributions improve the fit to 
the data on the antiproton to proton ratio for the narrow window $M\in [62.470,62.505] \text{ GeV}$ for the  tensor dark matter mass 
and $g_s \in [0.98,1.01]\times 10^{-3}$ for the Higgs portal coupling in the effective theory. These are precisely the range of 
values compatible with several experimental constraints, such as dark matter relic density, limits on the spin-independent dark 
matter-nucleon cross-section from XENON1T, indirect detection limits for the annihilation of dark matter into $\bar{b}b$, 
$\tau^+ \tau^-$, $\mu^+ \mu^-$ and $\gamma \gamma$, as well as with the gamma-ray excess at the Milky Way galactic center. 
\end{abstract}
\maketitle

\section{Introduction}

One of the major challenges in particle physics, astrophysics and cosmology is the elucidation of the nature of dark matter. There are 
several pieces of evidence for the existence of dark matter, such as galaxy rotation curves, relic density and those derived from the observation 
of the cosmic microwave background (two recent reviews on this subject can be found in \cite{Lin:2019uvt, Arbey:2021gdg}). For decades, a great effort 
has been put in the measurement of observables that help us to constrain the values of the mass and couplings (beyond gravity) 
of this type of matter that conforms $26\%$ of the energy content in the universe. Several experiments have been designed trying 
to detect signals of non-gravitational dark matter interactions in direct, indirect and collider searches. In particular, 
instruments such as PAMELA and AMS-02 have measured with good precision and sensitivity the antimatter cosmic-ray spectrum,  
providing information on the high-energy phenomena occurring in the Galaxy 
\cite{Adriani:2011cu,PhysRevLett.114.171103,PhysRevLett.115.211101,PhysRevLett.117.231102,PhysRevLett.117.091103} which may include antimatter 
produced in the annihilation of dark matter.

In the past decade, several reports indicate the existence of an excess of antiprotons in the $\sim10-20$ GeV region of the 
in the AMS-02 Collaboration data, and many studies have identified a consistency with an additional contribution 
from annihilating dark matter 
\cite{Bringmann:2014lpa,Cirelli:2014lwa,Hooper:2014ysa,Cui:2016ppb,Cuoco:2017rxb,Cholis:2019ejx}. It is interesting to note that 
most of these studies scan independently the values of $\langle \sigma v_r \rangle $ for the annihilation of dark matter into standard 
model particles, and the dark matter mass $M$, to conclude 
that it is possible to explain this excess with a dark matter particle of $M\sim60$ GeV, with an annihilation cross-section 
of the order of the thermal relic cross section ($\langle \sigma v_r \rangle \sim 10^{-26} cm^3 /seg$). A recent update on the antiproton to proton flux ratio has been published by the AMS Collaboration \cite{AGUILAR20211} and such high precision measurements requires careful consideration of the experimental and theoretical uncertainties.

Coincidentally, in Ref. \cite{Hernandez-Arellano:2019qgd}, the possibility that the gamma ray excess from the galactic center
claimed by several groups \cite{Hooper:2010mq,Boyarsky:2010dr,Hooper:2011ti,Abazajian:2012pn,Macias:2013vya,TheFermi-LAT:2017vmf,Gordon:2013vta} be due to the annihilation of tensor dark matter (TDM) into standard model particles was explored, finding that this is 
possible only if TDM has a mass in the range $M\in [62.470,62.505] \text{ GeV}$ and the coupling to the Higgs boson $g_{s}$, which 
naturally arises in the effective theory, satisfies $g_s \in [0.98,1.01]\times 10^{-3}$, which in turn yields $\langle \sigma v_r \rangle $ of the 
order of the thermal cross section for the annihilation into some standard model particles.

Tensor dark matter is an alternative description of dark matter fields introduced in \cite{Hernandez-Arellano:2018sen}, where dark 
matter fields transform in the $(1,0)\oplus(0,1)$ representation of the Homogenous Lorentz Group (HLG). The quantum field theory 
for these fields was constructed in \cite{Napsuciale:2015kua}, based on the general algorithm for the construction of a covariant basis 
of operators for fields transforming in the $(j,0)\oplus(0,j)$ representation outlined in \cite{Gomez-Avila:2013qaa}. Unlike the conventional 
description of fields transforming in this representation with antisymmetric tensor fields, charged tensor dark matter fields are 
described by a Dirac-like six-component spinor, satisfying a new equation of motion. In the effective field theory for the interactions of hidden tensor dark matter with standard model fields, the leading interacting terms for the simplest case of a $U(1)_D$ dark gauge group, turn out to be dimension-four and are given by \cite{Hernandez-Arellano:2018sen}
 \begin{equation}
\mathcal{L}_{int}=\bar{\psi}(g_{s}\mathbf{1}+ig_{p}\chi )\psi \tilde{\phi}%
\phi +g_{t}\bar{\psi}M_{\mu \nu }\psi B^{\mu \nu } + \mathcal{L}_{selfint},  \label{Leff}
\end{equation}%
where $g_{s}$, $g_{p}$ and $g_{t}$ are dimensionless constants. The first two yield parity-respecting and parity-violating \textit{Higgs portals} 
and the third couple the photon and $Z^{0}$ to higher multipoles of tensor dark matter and is therefore dubbed as \textit{spin portal}. After 
spontaneous symmetry breaking, this Lagrangian yields 
\begin{equation}
\mathcal{L}_{int}= \frac{1}{2}\bar\psi (g_{s} \mathbf{1}+ i g_{p}\chi) \psi
\left(H+ v\right)^2 + g_{t} C_{W} \bar\psi M_{\mu\nu} \psi F^{\mu\nu} - g_{t} S_{W} \bar\psi M_{\mu\nu} \psi Z^{\mu\nu} ,
\label{lag}
\end{equation}%
where $C_W$, $S_W$, $H$, $v$,  are the cosine and sine of the Weinberg angle, the Higgs field and 
the Higgs vacuum expectation value, and $F^{\mu\nu}, Z^{\mu\nu}$ stand for the stress tensors of the electromagnetic and 
$Z^0$ interactions.

In a first study of the constraints from available data it was shown that the measured relic density and invisible $Z$ width constrain the 
TDM mass to $M>43~GeV$. Also, XENON1T upper bounds \cite{Aprile:2017iyp}  for the spin independent dark matter-nucleon cross section constrain 
the spin portal coupling to $g_{t}<10^{-4}$ for dark matter masses of the order of hundreds of GeV and sensible values for $g_{s}\leq 10^{-1}$ 
are compatible with these upper bounds \cite{Hernandez-Arellano:2018sen}. Similarly, indirect detection upper bounds to the annihilation 
of dark matter into $\bar{b}b$, $\tau^+ \tau^-$ and $\mu^+ \mu^-$ are satisfied for a wide range of values of $M$. 

Recently, we studied the possibility that the excess in the gamma-ray spectrum from our galactic center be explained by the annihilation 
of tensor dark matter \cite{Hernandez-Arellano:2019qgd}. In a comprehensive study of the possible mechanisms, we found that the 
gamma ray excess can be accounted for by the annihilation of TDM only if $M\in [62.470,62.505] \text{ GeV}$. For a given $M$ in this
window, the measured relic density, $\Omega^{exp}_{DM} h^{2} =0.1186\pm0.0020 $ \cite{Ade:2015xua,Patrignani:2016xqp}, yields the corresponding value $g_{s}(M)$, which takes values in the range 
$g_s(M) \in [0.98,1.01]\times 10^{-3}$. These values of the TDM mass are at the Higgs 
resonance for non-relativistic dark matter, thus we refined our previous calculations to focus on the resonance region and to 
account for the resonant effects in the calculation 
of the relic density, finding that these values are compatible with the measured relic density, the XENON1T upper bounds on the 
dark matter-nucleon spin independent cross section, the upper bounds to the annihilation of dark matter into $\bar{b}b$, $\tau^+ \tau^-$ 
and $\mu^+ \mu^-$ from indirect detection experiments \cite{Bergstrom:2013jra,Aguilar:2013qda,Drlica-Wagner:2015xua,Fermi-LAT:2016uux} and also with the upper bounds for the annihilation into $\gamma \gamma$, from data from the FermiLAT \cite{Ackermann:2015lka} and HESS \cite{Abdallah:2018qtu} collaborations.

For the values of $M$ compatible with the gamma ray excess described above, we can have TDM annihilation into two fermions, 
which in turn travel through the interstellar medium (ISM), propagate and hadronize to generate other products, including antiprotons. 
This signal could potentially be sizable enough to be detected on Earth by any of the current anti-matter measuring experiments, 
including PAMELA and AMS-02. Annihilation into $\bar{b}b$ is expected to play a major role in the antiproton production, but there 
could be also contributions from the $\bar{c}c$ and $\tau^{+}\tau^{-}$ channels, The results obtained so far for TDM phenomenology
encouraged us to carry out the corresponding calculations whose results are presented in this paper.

Our work is organized as follows. In the next section we calculate the antiproton and proton cosmic-ray spectrum in the galaxy using 
standard methods and reproduce results in the literature for the production of antiprotons from standard sources. 
In section 3, we include the contribution from TDM annihilation into fermions. The resulting fits are compared to the results without 
TDM contributions to assess if there is a clear preference for TDM annihilation from the AMS-02. Our conclusions are given in section 4.

\section{Antiproton excess in the cosmic-ray spectrum}

A necessity for the study and observation for baryon asymmetry prompted the search and, eventually, detection of cosmic 
ray (CR) antiprotons \cite{Golden:1979bw,bogomolov1979stratospheric} during the 1970s. Following several measurements 
and model proposals of the generation of antiparticles in the Galaxy, it was concluded that CR antiprotons are produced after 
interactions between high-energy nuclei (cosmic ray primaries, i.e. those accelerated by remnants of supernova) and 
interestelar gas. These are called secondary antiprotons, and their energy spectrum is observed to have a maximum 
near $2~ GeV$ and then decrease at energies around tenths of $GeVs$ in a steeper form than that of protons. This results in a 
steep decrease in the antiproton-to-proton ratio at these energies. The antiproton flux is, thus, determined by CR propagation 
and interaction of nuclei with interestelar gas, processes which suffer from large uncertainties and that must be treated carefully in 
order to predict the flux with high precision.

The first source of systematic uncertainty comes from the CR propagation which involves several complicated processes 
such as diffusion, convection, re-acceleration and loss of energy. The parameters of choice to model such effects must account 
for secondary-to-primary nuclei ratios, among which are Boron-to-Carbon (B/C) and other nuclei \cite{Strong_1998}. There are 
many different ways to model these measurements, so the choice presents a difficult task and a great source of uncertainty.

A second source of uncertainty is the effect of solar modulation on the CR spectra \cite{osti_4797303}. When CR enter the 
Solar System, the solar magnetic field modifies their spectra. This mainly acts on the low energy part of the spectrum, but it 
is difficult to estimate it with precision since it requires the modeling of the solar wind and its effects, which change through time.

In this section we will make use of the parameter choice indicated by the first model in Ref. \cite{Cholis:2019ejx} to evaluate 
the proton and antiproton flux from cosmic rays.

\subsection{Modeling the antiproton and proton cosmic-ray spectrum in the galaxy}


We make use of the tool GALPROP\footnote{https://galprop.stanford.edu/} \footnote{https://galprop.stanford.edu/webrun/} \cite{2011CoPhC.182.1156V,2005AdSpR..35..156M,1998A&A...338L..75M}, 
which takes care of solving the transport equation to yield the local flux of the primary and secondary cosmic ray species, which can be written as:
\begin{equation}
    \frac{\partial \psi (\Vec{r},p,t)}{\partial t} = Q(\Vec{r},p,t)+\Vec{\nabla}\cdot (D_{xx} \Vec{\nabla}\psi - \Vec{V}\psi)+\frac{\partial}{\partial p} p^2 D_{pp} \frac{\partial}{\partial p} \frac{1}{p^2} \psi - \frac{\partial}{\partial p} \Big[ \dot{p} \psi - \frac{p}{3} (\Vec{\nabla}\cdot \Vec{V}) \psi \Big] - \frac{1}{\tau_f} \psi - \frac{1}{\tau_{r}}\psi ,
\end{equation}
where $\psi (\Vec{r},p,t)$ is the cosmic ray density per unit of particle momentum $p$ at position $\Vec{r}$. The source term, $Q(\Vec{r},p,t)$, includes the source injection spectrum (assumed to be radially distributed e.g. supernova remnants) and contribution from spallation \cite{Strong:2007nh}. Acceleration due to injection in general can be parametrized to the cosmic-ray rigidity $R$ by a power-law spectrum $dN/dR \sim R^{-\alpha}$, where $\alpha$ is called nucleus injection index. Generally, such spectrum can have breaks at rigidity $R_{br}$, with $\alpha=\alpha_1$ below this value and $\alpha=\alpha_2$ above it\cite{Ptuskin:2005ax}. 

$\Vec{V}$ is the convection velocity, while the last two terms involve momentum loss by fragmentation (with time scale $\tau_f$) and from radioactive decay (with time scale $\tau_r$).

The spatial diffusion coefficient is defined by
\begin{equation}
D_{xx} (R) = \beta D_0 (R / 4 \text{ GV})^\delta ,
\end{equation}
where $\delta$ is the diffusion index and $\beta \equiv v/c$. The momentum diffusion coefficient, $D_{pp}$, is inversely proportional to $D_{xx}$ and proportional to the squared Alfvén velocity $v_A$.

A study of the antiproton production cross section was done in Ref. \cite{diMauro:2014zea}, where it was found that different 
procedures led to equivalent results, yielding an uncertainty of 10-20\%. In order to account for this uncertainty, we employ 
the same approach as formulated in \cite{PhysRevD.95.123007}, where we multiply the antiproton flux prior solar modulation 
by an scaling factor in the form of an energy-dependent function, defined by
\begin{equation}
N_{CS} (k_{ISM}) = a + b \Big[ln \Big(\frac{k_{ISM}}{GeV} \Big)\Big] +  c \Big[ ln \Big( \frac{k_{ISM}}{GeV} \Big) \Big]^2 
+ d \Big[ ln \Big( \frac{k_{ISM}}{GeV} \Big) \Big]^3 ,
\end{equation}
where $k_{ISM}$ is the kinetic energy of the cosmic ray in the interstellar medium (ISM), before entering the Solar System. 
The fourth term of this scaling factor, proportional to $\Big[ ln \Big( \frac{k_{ISM}}{GeV} \Big) \Big]^3$, can be omitted and 
still reach a adequate fit, as we will show later. 

We must also consider the effects of solar modulation. The differential flux at Earth, $dN^\oplus / dE_{kin}$ in 
terms of $k_{ISM}$, is obtained as \cite{osti_4797303}
\begin{equation}
\frac{dN^\oplus}{dE_{kin}} (k_{ISM}) = \frac{(k_{ISM} - |Z|e \Phi (R) + m)^2 - m^2}{(k_{ISM} + m)^2 - m^2} 
 \frac{dN^{ISM}}{dk_{ISM}} (k_{ISM}) ,
\end{equation}
where ${dN^{ISM}}/{dk_{ISM}}$ is the differential flux prior to the effects of solar modulation, $E_{kin}$, $|Z|e$ and $m$ are 
the kinetic energy, charge and mass of the cosmic ray. To obtain the flux in terms of $E_{kin}$, we use the equivalence 
$E_{kin} = k_{ISM}-|Z|e \Phi(R)$. The modulation potential, $\Phi$, can be described as a function that depends on time, the 
rigidity ($R = \sqrt{k_{ISM}(k_{ISM}+2 m_p)}$) and charge of the cosmic ray prior entering the Solar System. We follow the 
analytic expression constructed in Ref. \cite{PhysRevD.93.043016}:
\begin{equation}
\Phi (R,t,q) = \phi_0 \Big(\frac{|B_{tot} (t)|}{4 \text{ nT}} \Big) + \phi_1 N' (q) H(-q A(t))
\Big( \frac{|B_{tot} (t)|}{4 \text{ nT}} \Big) 
\Big( \frac{1+ (R/R_0)^2}{\beta (R/R_0)^3} \Big) \Big( \frac{\alpha (t)}{\pi/2} \Big)^4 ,
\end{equation}
where $\beta$ is the velocity, $R_0 \equiv 0.5 \text{ GV}$ and $B_{tot}$ is the strength of the heliospheric magnetic field (HMF) 
at Earth, which has a polarity $A(t)$, $H$ denotes the Heaviside function and $\alpha$ is the tilt angle of the heliospheric current sheet. 
$N'(q) \neq 1$ when the HMF does not have a well-defined polarity.

We allow $\phi_0 \in [0.32,0.38]$ GV and $\phi_1 \in [0,16]$ GV in order to stand within the uncertainties for the modulation 
potential described in Ref. \cite{PhysRevD.93.043016}. The values of $B_{tot}$, $\alpha$ and $N' (q) H(-q A(t))$ averaged 
over six-month intervals during the observation period by AMS-02 can be found in Table II of \cite{PhysRevD.95.123007}. 
We take these values to evaluate the potential and the flux for each period and then take the average of the obtained values 
for the final result.

There is an additional parameter, the local ISM gas density normalization, taken as an energy-independent factor, $g_{ISM}$. 
In total, there are seven free parameters that will be used for the fit: $\phi_0$, $\phi_1$, $a$, $b$, $c$, $d$ and $g_{ISM}$. 
The flux ratio is defined as follows.
\begin{equation}
R_{\bar{p} / p} = \frac{\Phi_{\bar{p}}}{\Phi_p} = g_{ISM}  \frac{\frac{dN^\oplus_{\bar{p}}}{dE_{kin}}}{\frac{dN^\oplus_p}{dE_{kin}}}.
\end{equation}

\subsection{Results of the fit to the antiproton-proton ratio without dark matter}

 We tested a range of parameter configurations based on the three proposals made in \cite{Cholis:2019ejx}. We determined the parameters that directly influence the antiproton-to-proton ratio and found the configuration with the best fit to the AMS-02 data using iMinuit\cite{iminuit}. It is important to mention that these parameter configurations need to be tested against the experimental bounds set by the observed B/C ratio\cite{AGUILAR20211}. The configurations from \cite{Cholis:2019ejx} are consistent with this ratio, and we checked the fourth configuration finding very similar results. The values of the parameters for each configuration are shown in Table \ref{model1}.

\begin{table}[h!]
\centering
\begin{tabular}{|c|c|c|c|c|}
\hline
\textbf{Parameter} & \textbf{Mod I} & \textbf{Mod II} & \textbf{Mod III} & \textbf{Best Fit} \\ \hline
$\delta$ & 0.40 & 0.50 & 0.40 & 0.40 \\ \hline
 $z_L$ (kpc) & 5.6 & 6.0 & 3.0 & 6 \\ \hline
 DM Core Radius (kpc) & 0.0 & 20.0 & 0.0 & 20.0 \\ \hline
 DM local mass density ($GeV cm^{-3}$) & 0.0 & 0.30 & 0.0 & 0.30 \\ \hline
$D_0 $ ($cm^2 s^{-1}$) & $4.85\times 10^{28}$ & $3.10\times 10^{28}$ & $2.67\times 10^{28}$ & $4.85\times 10^{28}$ \\ \hline
 $v_A$ (km/s) & 24.0 & 23.0 & 22.0 & 24.0 \\ \hline
 $dv_c / d|z|$ (km/s/kpc) & 1.0 & 9.0 & 3.0 & 1.0 \\ \hline
${\alpha}_1$ & 1.88 & 1.88 & 1.87 & 1.87 \\ \hline
${\alpha}_2$ & 2.38 & 2.45 & 2.41 & 2.38 \\ \hline
\end{tabular}
\caption{Cosmic-ray injection and propagation model parameter configurations. Mod I, Mod II and Mod II were adopted from Ref. \cite{Cholis:2019ejx}. $z_L$ is the magnitude for the minimum ($-z_L$) and maximum ($+z_L$) height of the cylindrical surface centered at the Galactic Center at which the transport equation is solved. $dv_c / d|z|$ refers to the convection velocity gradient, assumed to be linear \cite{Ptuskin:2005ax}. }
\label{model1}
\end{table}

\begin{figure}[h!]
\center
\includegraphics[width=\textwidth]{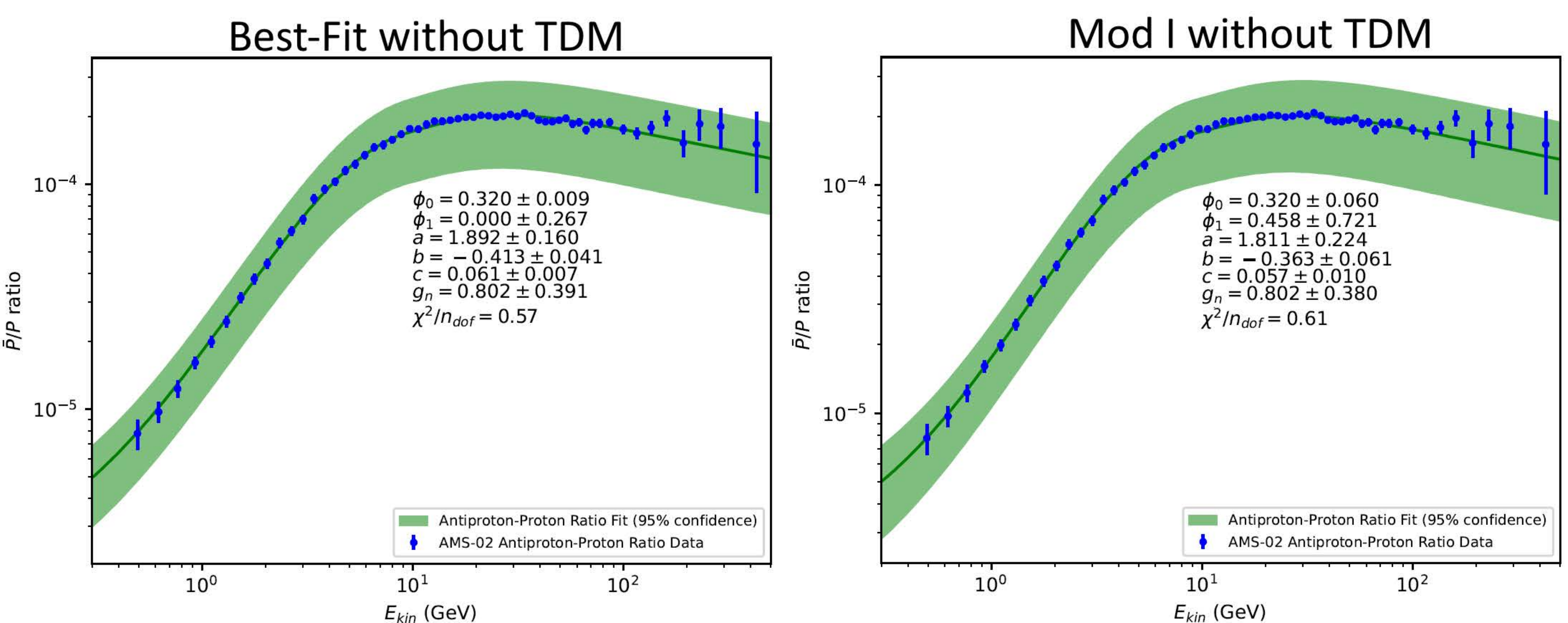}
\caption{Best fit to the AMS-02 data Antiproton-To-Proton ratio for our Best-Fit and Mod I parameter configurations. The shadowed band corresponds to the 95\% confidence level region.}
\label{antippratiofit}
\end{figure}

\begin{figure}[h!]
\center
\includegraphics[width=\textwidth]{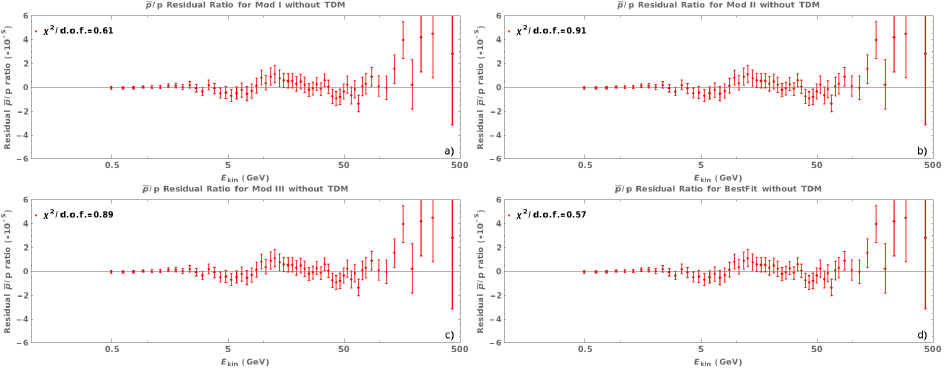}
\caption{Antiproton-to-proton ratio residuals from the AMS-02 data \cite{AGUILAR20211} for the parameter configurations: a) Mod I, b) Mod II, c) Mod III and d) our Best Fit. }
\label{AntiPPratioRes}
\end{figure}

The results for the fit to the AMS-02 antiproton-to-proton ratio data (Ref. \cite{AGUILAR20211}) from secondary antiproton productions without TDM contributions are shown in Fig. \ref{antippratiofit} for each of the configurations. A summary of the fitting parameters and the corresponding value of $\chi^2$ for each case is listed in Table \ref{resultswoDM}. The difference between the antiproton-to-proton ratio for each fit and the AMS-02 data is shown in Fig. \ref{AntiPPratioRes}. We can see that there is an excess in the residual around $\sim$10-20 GeV for all configurations, although less pronounced for the Best Fit. For all configurations a large difference is noticeable in the higher end of the spectrum. This discrepancy for higher energies is not within the purposes of this article, but it could be explained by secondary protons being accelerated by shocks from supernova remnants \cite{PhysRevD.95.123007}.

\begin{table}[h!]
\centering
\begin{tabular}{|c|c|c|c|c|}
\hline
\textbf{\begin{tabular}[c]{@{}c@{}}Fit \\ Parameters\end{tabular}} & \textbf{Mod I} & \textbf{Mod II} & \textbf{Mod III} & \textbf{Best Fit} \\ \hline
$\phi_0$/GV                                                           & 0.320          & 0.320           & 0.320            & 0.320             \\ \hline
$\phi_1$/GV                                                           & 0.458          & 1.243           & 3.010            & 0.000             \\ \hline
a                                                                  & 1.811          & 1.331           & 1.408            & 1.892             \\ \hline
b                                                                  & -0.363         & -0.058          & -0.126           & -0.413            \\ \hline
c                                                                  & 0.057          & 0.030           & 0.026            & 0.061             \\ \hline
$g_{ISM}$                                                              & 0.802          & 0.803           & 0.802            & 0.802             \\ \hline
$\chi^2$                                                           & 0.61           & 0.91            & 0.89             & 0.57              \\ \hline
\end{tabular}
\caption{Fit parameters to the AMS-02 antiproton-to-proton ratio data considering secondary antiprotons produced in the ISM without TDM contributions, for each of the configurations.}
\label{resultswoDM}
\end{table}

\section{Antiproton production from annihilating tensor dark matter}

Tensor dark matter can annihilate into fermion-anti-fermion through the Higgs portals and the spin portal in Eq. (\ref{lag}). The 
corresponding cross section is \cite{Hernandez-Arellano:2018sen}
\begin{align}
(\sigma v_r)_{\bar{f}f} (s)& =\frac{1}{ 144\pi M^4 \sqrt{s}} \frac{\sqrt{%
s-4m_{f}^{2}}}{(s-M^2)} \left[ \frac{ m_f^2 \left(s-4 m_f^2\right) \left( g_p^2 s
\left(s-4 M^2\right)+g_s^2 \left(6 M^4-4 M^2 s+s^2\right)\right)}{
\left(\left(s-M_H^2\right)^2 + \Gamma_H^2 M_H^2\right)} \right.  \notag \\
&\left. +\frac{2 g_{t}^2 M_Z^2 S_W^2 s\left(s-4 M^2 \right) \left(2
M^2+s\right) \left(2 \left(A_{f}^2-2 B_{f}^2\right) m_f^2+s
\left(A_{f}^2+B_{f}^2\right)\right)}{3 v^2 \left(\left(s-M_Z^2\right)^2 + \Gamma_Z^2 M_Z^2\right){}^2} \right.  \notag
\\
&\left. +\frac{32 C_W^2 Q_f^2 g_{t}^2 M_W^2 S_W^2 \left(s-4 M^2 \right)
\left(2 M^2+s\right) \left(2 m_f^2+s\right)}{3 v^2 s} \right.  \notag \\
&\left. -\frac{16 A_{f} C_W Q_f g_{t}^2 M_W M_Z S_W^2 \left(s-4 M^2 \right)
\left(2 M^2+s\right) \left(2 m_f^2+s\right)}{3 v^2 \left(\left(s-M_Z^2\right)^2 + \Gamma_Z^2 M_Z^2\right)} %
\right].
\label{sigmavffcomp}
\end{align}
where $m_f$, $Q_f$ correspond to the mass of the fermion and its charge in units of $e>0$, respectively, and
  \begin{equation}
  A_{f}= 2 T^{(3)}_{f} - 4 Q_{f} \sin^{2}\theta_{W}, \qquad B_{f}= - 2 T^{(3)}_{f}.
  \end{equation}
In the non-relativistic limit, only the parity-conserving Higgs portal contributions survive and the velocity averaged cross section reads
\begin{equation}
\langle\sigma v_{r} \rangle_{\bar{f}f}=\frac{N_{c}g^{2}_{s}m^{2}_{f} (M^{2}-m^{2}_{f} )^{\frac{3}{2}}} 
{12\pi M^{3}[(4M^{2}-M^{2}_{H})^{2} + M^{2}_{H} \Gamma^{2}_{H} ]} +{\cal{O}}(v^{2}_{r}),
\label{sigmavffeq}
\end{equation}
where $N_{c}=3$ for quarks and $N_{c}=1$ for leptons. Notice that $\langle\sigma v_{r} \rangle_{\bar{f}f}$ and the TDM mass $M$ 
are not independent parameters, thus the predictions are specific to the nature (space-time structure) of TDM. 
In order to keep consistency with the measured dark matter relic 
density, upper limits by XENON1T for the spin-independent dark matter-nucleon cross-section, indirect detection limits 
and the gamma-ray excess at the Milky Way galactic center, we consider $M\in [62.470,62.505] \text{ GeV}$ and 
$g_s \in [0.98,1.01]\times 10^{-3}$ in our calculations. 

\begin{figure}[h!]
\center
\includegraphics[scale=0.3]{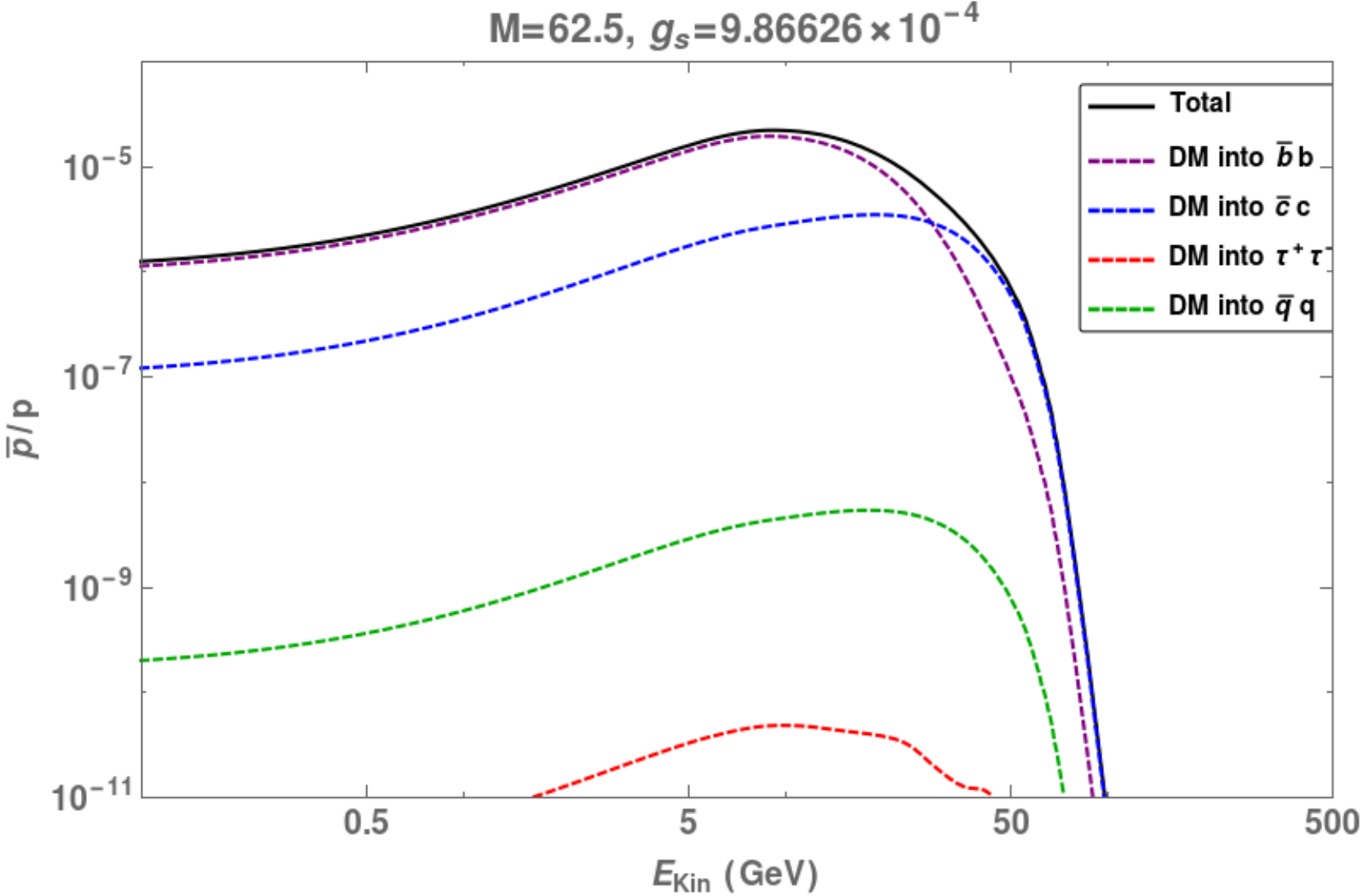}
\caption{Antiproton-to-proton flux ratio for antiprotons produced in the annihilation of TDM for $M=62.5$ GeV and 
$g_s = 9.866 \times 10^{-4}$, for the $\bar{b}b$, $\bar{c}c$, $\tau^+ \tau^-$ and light quark channels. }
\label{AntiPPratioTDM}
\end{figure}

We use of PPC4DMID code \cite{Cirelli_2011} to obtain the antiproton flux produced in the hadronization of quarks or the hadronic decay 
of the $\tau$ lepton produced in the annihilation of TDM, and its propagation in the ISM. The code uses Monte Carlo simulations in 
order to find the differential spectra of antiprotons, including electroweak corrections.  Although protons are also produced by the 
same mechanism, its contribution to the overall proton flux is very small and can be neglected. Only the antiproton production is large 
enough to be relevant in the calculation of the antiproton to proton flux ratio. The contributions of the relevant channels to the 
antiproton-proton ratio are shown in Fig. \ref{AntiPPratioTDM}.


\subsection{Results of the fit to the antiproton-proton ratio including tensor dark matter contributions}

\begin{figure}[h!]
\center
\includegraphics[width=\textwidth]{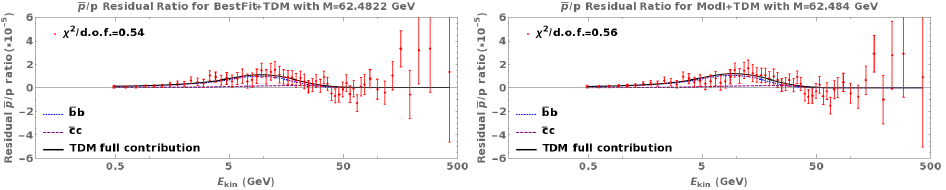}
\caption{Antiproton-to-proton ratio residual from the AMS-02 data from Ref. \cite{AGUILAR20211} and the best fit contribution from TDM annihilation, for the Mod I and our Best-Fit parameter configurations. We show the most relevant contributions, from $\bar{b}b$ and $\bar{c}c$ pairs, with the TDM mass value of M=62.484 GeV and M=62.4822 GeV for the Mod I and our Best-Fit configurations, respectively.}
\label{APPratioBestfitDMwD}
\end{figure}

\begin{figure}[h!]
\center
\includegraphics[width=\textwidth]{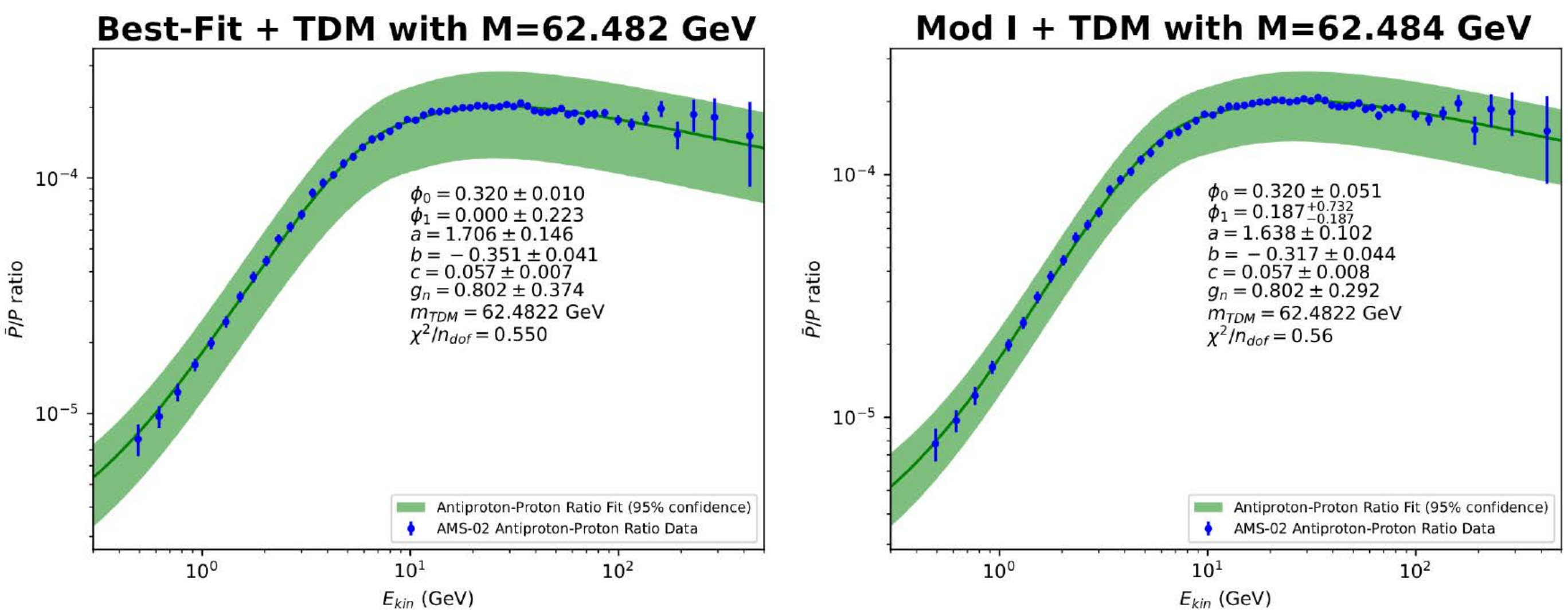}
\caption{Antiproton-to-proton ratio fits from the AMS-02 data from Ref. \cite{AGUILAR20211} including the best-fit contributions from TDM annihilation into fermion pairs, for the Mod I and our Best-Fit parameter configurations. The shadowed region corresponds to the 95\% confidence-level.}
\label{APPratioMinMassTDM}
\end{figure}

\begin{figure}[h!]
\center
\includegraphics[width=\textwidth]{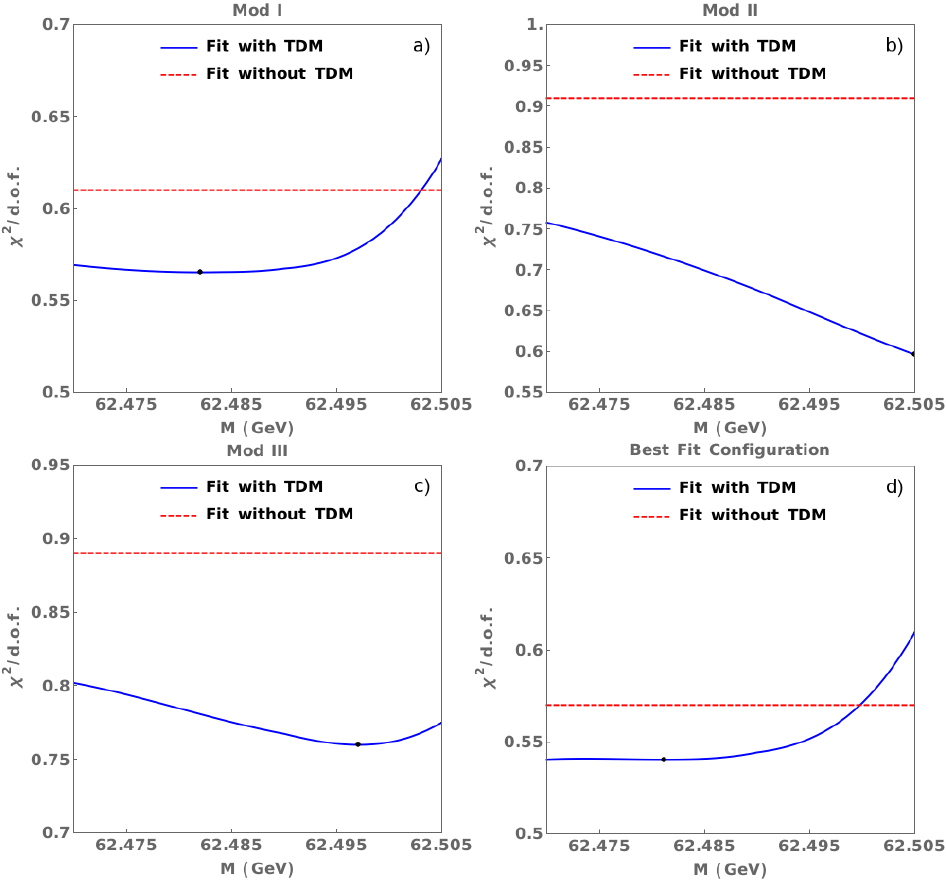}
\caption{We perform a fit for different values of the DM mass and we obtain the $\chi^2$ value for a) Mod I, b) Mod II, c) Mod III and d) our Best-Fit. We find that the fit is improved for almost all values of our mass interval, except for a small region near the resonance in the case of Mod I and our Best-Fit.}
\label{chisqrfit}
\end{figure}

We chose five different TDM masses, $M=62.505,62.500, 62.490,62.480\text{ and }62.470\text{ GeV}$, and for each value we performed the fit with the contribution from TDM annihilation into $\bar{f}f$ contributions. Table \ref{resultswDMwD} lists the parameters obtained in the fit. The residuals of the fit to the AMS-02 data, including the contributions from TDM annihilation into $\bar{f}f$ to the antiproton-to-proton ratio in the Mod I and Best-Fit configurations are shown in Fig. \ref{APPratioBestfitDMwD} for the value of M that produces the best fit. In Fig. \ref{chisqrfit} we plot the $\chi^2$ value of the fit including TDM and compare it to the value for the best fit without dark matter obtained in the previous section. Notice that the fit improves for almost all the range of mass values discussed, except for a small region near the resonance in the case of Mod I and our Best-Fit. In Fig. \ref{APPratioMinMassTDM} we plot the fits to the antiproton-to-proton ratio along with the 95\% confidence band for the Mod I and our Best-Fit configurations with an added contribution from TDM annihilation with the value of mass that yielded the best fit in each case.

\begin{table}[h]
\centering
\small
\begin{tabular}{|cccccccc|}
\hline
\multicolumn{8}{|c|}{\textbf{Mod I}}                                                                                                                                                                                                                                                         \\ \hline
\multicolumn{1}{|c|}{\textbf{M (GeV)}} & \multicolumn{1}{c|}{\textbf{$\phi_0$}} & \multicolumn{1}{c|}{\textbf{$\phi_1$}} & \multicolumn{1}{c|}{\textbf{a}} & \multicolumn{1}{c|}{\textbf{b}} & \multicolumn{1}{c|}{\textbf{c}} & \multicolumn{1}{c|}{\textbf{$g_{ISM}$}} & \textbf{$\chi^2$} \\ \hline
\multicolumn{1}{|c|}{62.505}           & \multicolumn{1}{c|}{0.320}             & \multicolumn{1}{c|}{0.000}             & \multicolumn{1}{c|}{1.466}      & \multicolumn{1}{c|}{-0.278}     & \multicolumn{1}{c|}{0.057}      & \multicolumn{1}{c|}{0.803}              & 0.63              \\ \hline
\multicolumn{1}{|c|}{62.500}           & \multicolumn{1}{c|}{0.320}             & \multicolumn{1}{c|}{0.000}             & \multicolumn{1}{c|}{1.527}      & \multicolumn{1}{c|}{-0.293}     & \multicolumn{1}{c|}{0.057}      & \multicolumn{1}{c|}{0.803}              & 0.59              \\ \hline
\multicolumn{1}{|c|}{62.490}           & \multicolumn{1}{c|}{0.320}             & \multicolumn{1}{c|}{0.123}             & \multicolumn{1}{c|}{1.606}      & \multicolumn{1}{c|}{-0.310}     & \multicolumn{1}{c|}{0.057}      & \multicolumn{1}{c|}{0.803}              & 0.57              \\ \hline
\multicolumn{1}{|c|}{62.480}           & \multicolumn{1}{c|}{0.320}             & \multicolumn{1}{c|}{0.215}             & \multicolumn{1}{c|}{1.656}      & \multicolumn{1}{c|}{-0.322}     & \multicolumn{1}{c|}{0.057}      & \multicolumn{1}{c|}{0.802}              & 0.56              \\ \hline
\multicolumn{1}{|c|}{62.470}           & \multicolumn{1}{c|}{0.320}             & \multicolumn{1}{c|}{0.270}             & \multicolumn{1}{c|}{1.689}      & \multicolumn{1}{c|}{-0.330}     & \multicolumn{1}{c|}{0.057}      & \multicolumn{1}{c|}{0.802}              & 0.57              \\ \hline
\multicolumn{8}{|c|}{\textbf{Mod II}}                                                                                                                                                                                                                                                        \\ \hline
\multicolumn{1}{|c|}{\textbf{M (GeV)}} & \multicolumn{1}{c|}{\textbf{$\phi_0$}} & \multicolumn{1}{c|}{\textbf{$\phi_1$}} & \multicolumn{1}{c|}{\textbf{a}} & \multicolumn{1}{c|}{\textbf{b}} & \multicolumn{1}{c|}{\textbf{c}} & \multicolumn{1}{c|}{\textbf{$g_{ISM}$}} & \textbf{$\chi^2$} \\ \hline
\multicolumn{1}{|c|}{62.505}           & \multicolumn{1}{c|}{0.320}             & \multicolumn{1}{c|}{0.695}             & \multicolumn{1}{c|}{1.163}      & \multicolumn{1}{c|}{-0.093}     & \multicolumn{1}{c|}{0.048}      & \multicolumn{1}{c|}{0.804}              & 0.60              \\ \hline
\multicolumn{1}{|c|}{62.500}           & \multicolumn{1}{c|}{0.320}             & \multicolumn{1}{c|}{0.839}             & \multicolumn{1}{c|}{1.189}      & \multicolumn{1}{c|}{-0.083}     & \multicolumn{1}{c|}{0.045}      & \multicolumn{1}{c|}{0.804}              & 0.62              \\ \hline
\multicolumn{1}{|c|}{62.490}           & \multicolumn{1}{c|}{0.320}             & \multicolumn{1}{c|}{0.966}             & \multicolumn{1}{c|}{1.227}      & \multicolumn{1}{c|}{-0.073}     & \multicolumn{1}{c|}{0.040}      & \multicolumn{1}{c|}{0.803}              & 0.68              \\ \hline
\multicolumn{1}{|c|}{62.480}           & \multicolumn{1}{c|}{0.320}             & \multicolumn{1}{c|}{1.074}             & \multicolumn{1}{c|}{1.252}      & \multicolumn{1}{c|}{-0.067}     & \multicolumn{1}{c|}{0.037}      & \multicolumn{1}{c|}{0.803}              & 0.72              \\ \hline
\multicolumn{1}{|c|}{62.470}           & \multicolumn{1}{c|}{0.320}             & \multicolumn{1}{c|}{1.119}             & \multicolumn{1}{c|}{1.268}      & \multicolumn{1}{c|}{-0.065}     & \multicolumn{1}{c|}{0.036}      & \multicolumn{1}{c|}{0.803}              & 0.76              \\ \hline
\multicolumn{8}{|c|}{\textbf{Mod III}}                                                                                                                                                                                                                                                       \\ \hline
\multicolumn{1}{|c|}{\textbf{M (GeV)}} & \multicolumn{1}{c|}{\textbf{$\phi_0$}} & \multicolumn{1}{c|}{\textbf{$\phi_1$}} & \multicolumn{1}{c|}{\textbf{a}} & \multicolumn{1}{c|}{\textbf{b}} & \multicolumn{1}{c|}{\textbf{c}} & \multicolumn{1}{c|}{\textbf{$g_{ISM}$}} & \textbf{$\chi^2$} \\ \hline
\multicolumn{1}{|c|}{62.505}           & \multicolumn{1}{c|}{0.320}             & \multicolumn{1}{c|}{2.916}             & \multicolumn{1}{c|}{1.137}      & \multicolumn{1}{c|}{-0.085}     & \multicolumn{1}{c|}{0.032}      & \multicolumn{1}{c|}{0.804}              & 0.78              \\ \hline
\multicolumn{1}{|c|}{62.500}           & \multicolumn{1}{c|}{0.320}             & \multicolumn{1}{c|}{2.958}             & \multicolumn{1}{c|}{1.180}      & \multicolumn{1}{c|}{-0.088}     & \multicolumn{1}{c|}{0.030}      & \multicolumn{1}{c|}{0.803}              & 0.76              \\ \hline
\multicolumn{1}{|c|}{62.490}           & \multicolumn{1}{c|}{0.320}             & \multicolumn{1}{c|}{2.988}             & \multicolumn{1}{c|}{1.241}      & \multicolumn{1}{c|}{-0.095}     & \multicolumn{1}{c|}{0.028}      & \multicolumn{1}{c|}{0.803}              & 0.77              \\ \hline
\multicolumn{1}{|c|}{62.480}           & \multicolumn{1}{c|}{0.320}             & \multicolumn{1}{c|}{3.000}             & \multicolumn{1}{c|}{1.281}      & \multicolumn{1}{c|}{-0.101}     & \multicolumn{1}{c|}{0.028}      & \multicolumn{1}{c|}{0.803}              & 0.78              \\ \hline
\multicolumn{1}{|c|}{62.470}           & \multicolumn{1}{c|}{0.320}             & \multicolumn{1}{c|}{3.003}             & \multicolumn{1}{c|}{1.308}      & \multicolumn{1}{c|}{-0.106}     & \multicolumn{1}{c|}{0.027}      & \multicolumn{1}{c|}{0.803}              & 0.80              \\ \hline
\multicolumn{8}{|c|}{\textbf{Best Fit}}                                                                                                                                                                                                                                                      \\ \hline
\multicolumn{1}{|c|}{\textbf{M (GeV)}} & \multicolumn{1}{c|}{\textbf{$\phi_0$}} & \multicolumn{1}{c|}{\textbf{$\phi_1$}} & \multicolumn{1}{c|}{\textbf{a}} & \multicolumn{1}{c|}{\textbf{b}} & \multicolumn{1}{c|}{\textbf{c}} & \multicolumn{1}{c|}{\textbf{$g_{ISM}$}} & \textbf{$\chi^2$} \\ \hline
\multicolumn{1}{|c|}{62.505}           & \multicolumn{1}{c|}{0.320}             & \multicolumn{1}{c|}{0.000}             & \multicolumn{1}{c|}{1.507}      & \multicolumn{1}{c|}{-0.293}     & \multicolumn{1}{c|}{0.054}      & \multicolumn{1}{c|}{0.803}              & 0.61              \\ \hline
\multicolumn{1}{|c|}{62.500}           & \multicolumn{1}{c|}{0.320}             & \multicolumn{1}{c|}{0.000}             & \multicolumn{1}{c|}{1.571}      & \multicolumn{1}{c|}{-0.311}     & \multicolumn{1}{c|}{0.055}      & \multicolumn{1}{c|}{0.803}              & 0.57              \\ \hline
\multicolumn{1}{|c|}{62.490}           & \multicolumn{1}{c|}{0.320}             & \multicolumn{1}{c|}{0.000}             & \multicolumn{1}{c|}{1.660}      & \multicolumn{1}{c|}{-0.337}     & \multicolumn{1}{c|}{0.056}      & \multicolumn{1}{c|}{0.802}              & 0.54              \\ \hline
\multicolumn{1}{|c|}{62.480}           & \multicolumn{1}{c|}{0.320}             & \multicolumn{1}{c|}{0.000}             & \multicolumn{1}{c|}{1.716}      & \multicolumn{1}{c|}{-0.355}     & \multicolumn{1}{c|}{0.057}      & \multicolumn{1}{c|}{0.802}              & 0.54              \\ \hline
\multicolumn{1}{|c|}{62.470}           & \multicolumn{1}{c|}{0.320}             & \multicolumn{1}{c|}{0.000}             & \multicolumn{1}{c|}{1.754}      & \multicolumn{1}{c|}{-0.367}     & \multicolumn{1}{c|}{0.058}      & \multicolumn{1}{c|}{0.802}              & 0.54              \\ \hline
\end{tabular}
\caption{Fit parameters to the AMS-02 antiproton-to-proton ratio data considering secondary antiprotons produced in the ISM and from TDM annihilation for Mod I, Mod II, Mod III and our Best-Fit.}
\label{resultswDMwD}

\end{table}


%


 \section{Conclusions}

In this work we study the production of antiprotons in the annihilation of tensor dark matter and the possibility that it explains the 
cosmic-ray antiprotons excess in the Galaxy and the antiproton-to-proton ratio in the AMS-02 data claimed by 
several groups. First we perform fits to the AMS-02 data without including dark matter, using the GALPROP tool to solve 
the transport equation in order to evaluate the primary and secondary proton and antiproton flux produced in the interestellar 
medium for four different parameter configurations. We employ a scaling factor as it was done in \cite{PhysRevD.95.123007}, which involves the 
uncertainty regarding the antiproton production cross-section. The fits are improved by this scaling factor and our results 
point to a possible excess of antiprotons in the $\sim10-20$ GeV region of the spectrum. 

Finally, we use the PPC4DMID code to calculate the flux of antiprotons produced in the hadronization and propagation of quarks or in 
the hadronic decays of leptons produced in the annihilation of tensor dark matter. The corresponding velocity averaged cross sections depend 
only on the coupling $g_{s}$ of the parity-conserving Higgs portal in the effective theory and on the TDM mass $M$. The values of these 
parameters are constrained by the measured relic density, the gamma ray excess from the galactic center and the upper bounds for the 
dark matter-nucleon spin independent cross section measured by XENON to the narrow windows  $g_s \in [0.98,1.01]\times 10^{-3}$ and 
$M \in [62.470,62.505] \text{ GeV}$. In particular the measured relic density correlate the values of these parameters such that 
$g_{s}=g_{s}(M)$ and we have actually only one independent parameter. These range of values are also consistent with the upper 
bounds for the annihilation of dark matter into $\bar{b}b$, $\tau^{+}\tau{-}$, $\mu^{+}\mu^{-}$ and $\gamma\gamma$ from indirect 
detection searches in gamma ray physics \cite{Hernandez-Arellano:2019qgd}. We find that the fit to the antiproton-to-proton ratio is 
improved when we include the contributions from TDM for almost all the values of $M \in [62.470,62.505] \text{ GeV}$ and the corresponding $g_{s}(M)$, except for a small region near the resonance for the Mod I and our Best-Fit configurations.

It is important to note that fitting to the antiproton flux poses a challenge particularly because of the large uncertainties while modeling the production and subsequent propagation of cosmic rays, and in order to make robust conclusions about dark matter one must act cautiously \cite{Calore:2022stf,Balan_2023}. By performing our analysis, we take into account the sources of uncertainty in a conservative way, finding that the addition of TDM annihilation indeed improves the fit in all cases. Some studies that focus on the systematic uncertainties of the data concluded that the presence of an antiproton excess from cosmic rays detected by the AMS-02 experiment are a product of secondary processes \cite{Boudaud:2019efq,Lv:2023gdt}, and do not have a main primary origin such as production from dark matter annihilation. However, the antimatter cosmic-ray spectrum, such as the antiproton-to-proton ratio, remains an important observable for constraining the parameters of dark matter models. If dark matter, through a certain process, were to produce antiprotons, it must be consistent with the observations. In the case of TDM, its contribution to the antiproton-to-proton ratio improves the fit to the data, affirming its consistency with the AMS-02 measurements, as well as with bounds from both indirect and direct detection searches.

\section{Acknowledgements}

One of us (R. Ramírez-Guzmán) acknowledges the scholarship provided by CONAHCYT-Mexico in the pursuit of his PhD.

\bibliographystyle{prsty}
\bibliography{dm2}

\begin{thebibliography}{10}

\bibitem{Lin:2019uvt}
T. Lin, PoS {\bf 333},  009  (2019).

\bibitem{Arbey:2021gdg}
A. Arbey and F. Mahmoudi, Prog. Part. Nucl. Phys. {\bf 119},  103865  (2021).

\bibitem{Adriani:2011cu}
O. Adriani {\it et~al.}, Science {\bf 332},  69  (2011).

\bibitem{PhysRevLett.114.171103}
M. Aguilar {\it et~al.}, Phys. Rev. Lett. {\bf 114},  171103  (2015).

\bibitem{PhysRevLett.115.211101}
M. Aguilar {\it et~al.}, Phys. Rev. Lett. {\bf 115},  211101  (2015).

\bibitem{PhysRevLett.117.231102}
M. Aguilar {\it et~al.}, Phys. Rev. Lett. {\bf 117},  231102  (2016).

\bibitem{PhysRevLett.117.091103}
M. Aguilar {\it et~al.}, Phys. Rev. Lett. {\bf 117},  091103  (2016).

\bibitem{Bringmann:2014lpa}
T. Bringmann, M. Vollmann, and C. Weniger, Phys. Rev. D {\bf 90},  123001
  (2014).

\bibitem{Cirelli:2014lwa}
M. Cirelli {\it et~al.}, JCAP {\bf 12},  045  (2014).

\bibitem{Hooper:2014ysa}
D. Hooper, T. Linden, and P. Mertsch, JCAP {\bf 03},  021  (2015).

\bibitem{Cui:2016ppb}
M.-Y. Cui, Q. Yuan, Y.-L.~S. Tsai, and Y.-Z. Fan, Phys. Rev. Lett. {\bf 118},
  191101  (2017).

\bibitem{Cuoco:2017rxb}
A. Cuoco, J. Heisig, M. Korsmeier, and M. Kr\"amer, JCAP {\bf 10},  053
  (2017).

\bibitem{Cholis:2019ejx}
I. Cholis, T. Linden, and D. Hooper, Phys. Rev. D {\bf 99},  103026  (2019).

\bibitem{AGUILAR20211}
M. Aguilar {\it et~al.}, Physics Reports {\bf 894},  1  (2021), the Alpha
  Magnetic Spectrometer (AMS) on the International Space Station: Part II -
  Results from the First Seven Years.

\bibitem{Hernandez-Arellano:2019qgd}
H. Hern\'andez-Arellano, M. Napsuciale, and S. Rodr\'\i{}guez, JHEP {\bf 08},
  106  (2020).

\bibitem{Hooper:2010mq}
D. Hooper and L. Goodenough, Phys. Lett. {\bf B697},  412  (2011).

\bibitem{Boyarsky:2010dr}
A. Boyarsky, D. Malyshev, and O. Ruchayskiy, Phys. Lett. {\bf B705},  165
  (2011).

\bibitem{Hooper:2011ti}
D. Hooper and T. Linden, Phys. Rev. {\bf D84},  123005  (2011).

\bibitem{Abazajian:2012pn}
K.~N. Abazajian and M. Kaplinghat, Phys. Rev. {\bf D86},  083511  (2012),
  [Erratum: Phys. Rev.D87,129902(2013)].

\bibitem{Macias:2013vya}
O. Macias and C. Gordon, Phys. Rev. {\bf D89},  063515  (2014).

\bibitem{TheFermi-LAT:2017vmf}
M. Ackermann {\it et~al.}, Astrophys. J. {\bf 840},  43  (2017).

\bibitem{Gordon:2013vta}
C. Gordon and O. Macias, Phys. Rev. {\bf D88},  083521  (2013), [Erratum: Phys.
  Rev.D89,no.4,049901(2014)].

\bibitem{Hernandez-Arellano:2018sen}
H. Hern\'andez-Arellano, M. Napsuciale, and S. Rodr\'\i{}guez, Phys. Rev. D
  {\bf 98},  015001  (2018).

\bibitem{Napsuciale:2015kua}
M. Napsuciale, S. Rodr\'\i{}guez, R. Ferro-Hern\'andez, and S. G\'omez-\'Avila,
  Phys. Rev. D {\bf 93},  076003  (2016).

\bibitem{Gomez-Avila:2013qaa}
S. G\'omez-\'Avila and M. Napsuciale, Phys. Rev. D {\bf 88},  096012  (2013).

\bibitem{Aprile:2017iyp}
E. Aprile {\it et~al.}, Phys. Rev. Lett. {\bf 119},  181301  (2017).

\bibitem{Ade:2015xua}
P.~A.~R. Ade {\it et~al.}, Astron. Astrophys. {\bf 594},  A13  (2016).

\bibitem{Patrignani:2016xqp}
C. Patrignani {\it et~al.}, Chin. Phys. {\bf C40},  100001  (2016 and 2017
  update).

\bibitem{Bergstrom:2013jra}
L. Bergstrom {\it et~al.}, Phys. Rev. Lett. {\bf 111},  171101  (2013).

\bibitem{Aguilar:2013qda}
M. Aguilar {\it et~al.}, Phys. Rev. Lett. {\bf 110},  141102  (2013).

\bibitem{Drlica-Wagner:2015xua}
A. Drlica-Wagner {\it et~al.}, Astrophys. J. {\bf 809},  L4  (2015).

\bibitem{Fermi-LAT:2016uux}
A. Albert {\it et~al.}, Astrophys. J. {\bf 834},  110  (2017).

\bibitem{Ackermann:2015lka}
M. Ackermann {\it et~al.}, Phys. Rev. {\bf D91},  122002  (2015).

\bibitem{Abdallah:2018qtu}
H. Abdallah {\it et~al.}, Phys. Rev. Lett. {\bf 120},  201101  (2018).

\bibitem{Golden:1979bw}
R.~L. Golden {\it et~al.}, Phys. Rev. Lett. {\bf 43},  1196  (1979).

\bibitem{bogomolov1979stratospheric}
E. Bogomolov {\it et~al.},  in {\em International Cosmic Ray Conference}
  (PUBLISHER, ADDRESS, 1979), Vol.~1, p.\ 330.

\bibitem{Strong_1998}
A.~W. Strong and I.~V. Moskalenko, The Astrophysical Journal {\bf 509},  212
  (1998).

\bibitem{osti_4797303}
L.~J. Gleeson and W.~I. Axford, Astrophys. J., 154: 1011-26(Dec. 1968).  .

\bibitem{2011CoPhC.182.1156V}
A.~E. {Vladimirov} {\it et~al.}, Computer Physics Communications {\bf 182},
  1156  (2011).

\bibitem{2005AdSpR..35..156M}
I.~V. {Moskalenko}, A.~W. {Strong}, J.~F. {Ormes}, and S.~G. {Mashnik},
  Advances in Space Research {\bf 35},  156  (2005).

\bibitem{1998A&A...338L..75M}
I.~V. Moskalenko, A.~W. Strong, and O. Reimer, Astron. Astrophys. {\bf 338},
  L75  (1998).

\bibitem{Strong:2007nh}
A.~W. Strong, I.~V. Moskalenko, and V.~S. Ptuskin, Ann. Rev. Nucl. Part. Sci.
  {\bf 57},  285  (2007).

\bibitem{Ptuskin:2005ax}
V.~S. Ptuskin {\it et~al.}, Astrophys. J. {\bf 642},  902  (2006).

\bibitem{diMauro:2014zea}
M. di~Mauro, F. Donato, A. Goudelis, and P.~D. Serpico, Phys. Rev. D {\bf 90},
  085017  (2014), [Erratum: Phys.Rev.D 98, 049901 (2018)].

\bibitem{PhysRevD.95.123007}
I. Cholis, D. Hooper, and T. Linden, Phys. Rev. D {\bf 95},  123007  (2017).

\bibitem{PhysRevD.93.043016}
I. Cholis, D. Hooper, and T. Linden, Phys. Rev. D {\bf 93},  043016  (2016).

\bibitem{iminuit}
H. Dembinski and P.~O. et~al.,   (2020).

\bibitem{Cirelli_2011}
M. Cirelli {\it et~al.}, Journal of Cosmology and Astroparticle Physics {\bf
  2011},  051  (2011).

\bibitem{Calore:2022stf}
F. Calore {\it et~al.}, SciPost Phys. {\bf 12},  163  (2022).

\bibitem{Balan_2023}
S. Balan {\it et~al.}, Journal of Cosmology and Astroparticle Physics {\bf
  2023},  052  (2023).

\bibitem{Boudaud:2019efq}
M. Boudaud {\it et~al.}, Phys. Rev. Res. {\bf 2},  023022  (2020).

\bibitem{Lv:2023gdt}
X.-J. Lv {\it et~al.}, Phys. Rev. D {\bf 109},  043006  (2024).

\end{thebibliography}

\end{document}